\documentclass[a4paper,11pt]{article}
\usepackage{pos}

\usepackage[utf8]{inputenc}
\usepackage{amsmath}
\usepackage{amsfonts}
\usepackage{graphicx}
\usepackage{hyperref}
\usepackage{braket}
\usepackage{multirow}
\usepackage{enumitem}
\usepackage{bbm}
\usepackage{units}
\usepackage{tikz}
\usepackage{subfloat}
\usetikzlibrary{fadings,shadows,positioning,calc,matrix,shapes,decorations.markings,shapes.arrows}

\graphicspath{{plots/}}

\definecolor{red}{rgb}{.9,0,0}
\definecolor{blue}{rgb}{0,0,.75}
\definecolor{green}{rgb}{0,.6,0}
\definecolor{magenta}{cmyk}{0,1,0,0}
\definecolor{myyellow}{rgb}{.98,.84,.37}

\newcommand{\olsi}[1]{\,\overline{\!{#1}}} 

\newcommand{\f}{f}
\newcommand{\cS}{\widetilde{\cal T}}  
\newcommand{\cK}{\olsi K}	
\newcommand{\cR}{{\cal T}}  
\newcommand{\cG}{{\cal K}}	
\newcommand{\cT}{\overline T}	
\newcommand{\cc}{\tilde c}  
\newcommand{\cj}{\tilde j}  
\newcommand{\cf}{\tilde f}	
\newcommand{\xx}{(x,x+\hat1)}
\newcommand{\X}{\cx}
\newcommand{\cx}{X}

\newcommand{\ind}[1]{[#1_{x}]}
\newcommand{\indx}[2]{[#1_{#2}]}

\renewcommand{\j}{j}

\newcommand{\SU}{\text{SU}}

\def\sl3c{\text{SL}(3,\mathbb{C})}
\def\su3{\text{SU(3)}}

\newcommand{\J}{\boldsymbol{j}}

\newcommand{\logZ}{\log Z}
\newcommand{\revprod}[1]{{\coprod_{#1}}}
\newcommand{\inlinerevprod}[1]{{\coprod_{#1}}}

\title{Grassmann tensor-network method for strong-coupling QCD}
 
\author*[a]{Jacques Bloch}
\author[a,b]{Robert Lohmayer}

\affiliation[a]{Institute for Theoretical Physics, University of Regensburg, Regensburg, Germany}
\affiliation[b]{Leibniz Institute for Immunotherapy, Regensburg, Germany}

\emailAdd{jacques.bloch@ur.de}
\emailAdd{robert.lohmayer@ur.de}

\abstract{We present a tensor-network method for strong-coupling QCD with staggered quarks at nonzero chemical potential. After integrating out the gauge fields at infinite coupling, the partition function can be written as a full contraction of a tensor network consisting of coupled local numeric and Grassmann tensors. To evaluate the partition function and to compute observables, we develop a Grassmann higher-order tensor renormalization group method, specifically tailored for this model. We apply the method to the two-dimensional case and validate it by comparing results for the partition function, the chiral condensate and the baryon density with exact analytical expressions on small lattices up to volumes of $4\times4$. For larger two-dimensional volumes, we present tensor results for the chiral condensate as a function of the mass and volume, and observe that the chiral symmetry is not broken dynamically in two dimensions. Furthermore, our results for the number density as a function of the chemical potential hint at a first-order phase transition. Finally, we present some preliminary tensor results for three-dimensional strong-coupling QCD.}

\FullConference{%
  The 39th International Symposium on Lattice Field Theory (Lattice2022),\\
  8-13 August, 2022 \\
  Bonn, Germany 
}

\begin{document}
\maketitle
 
\section{Introduction}

Simulations of classical or quantum systems in thermal equilibrium are typically performed using Monte Carlo (MC) methods, which are based on stochastic sampling. As an alternative, the partition function and the derived thermodynamic observables can also be computed using tensor methods when the partition function can be rewritten as a fully contracted tensor network \cite{Levin:2006jai,Xie_2012}. This has already been done successfully for various classical and quantum statistical systems in two, three and four dimensions \cite{Meurice:2020pxc}. For fermionic systems, where the partition function can no longer be written as a tensor network when the fermionic degrees of freedom are integrated out, Grassmann tensor-network methods were recently developed. In these proceedings we present such a Grassmann tensor-network method for strong-coupling QCD (sQCD) in $d\geq2$ dimensions. Although the original tensor-network methods, like the higher-order tensor renormalization group (HOTRG) method, are quasi independent of the model to which it is applied, this is different with the Grassmann tensor-network methods like GTRG \cite{Shimizu:2014uva,Takeda:2014vwa} and GHOTRG \cite{Sakai:2017jwp}, where the precise fermionic content of the theory influences the method substantially, such that the latter hast to be tailored specifically to the theory being considered. A detailed explanation of the method and results for the two-dimensional case were already published in Ref.~\cite{Bloch:2022vqz}.

Although the numeric complexity of tensor methods is usually higher that that of MC methods, it has two important advantages: It can be used to compute results for theories with a complex action, which leads to the sign problem in MC methods, and the method scales logarithmically in the volume, which allows for the investigation of the theory in very large volumes. Note that tensor-network methods are deterministic and are based on higher-order singular value decompositions (HOSVD) \cite{DeLathauwer2000}.

\section{Strong-coupling QCD}

\subsection{Partition function}

The partition function of sQCD, where $\beta=0$, is
\begin{align}
Z = \int \left[\prod_x d\psi_x d\bar\psi_x \prod_\nu dU_{x,\nu} \right]e^{S_F} 
\end{align}
with $d$-dimensional fermion action for one staggered quark with mass $m$ and chemical potential $\mu$,
\begin{align}
S_F = \sum_x \left\{
 \sum_{\nu=1}^d \eta_{x,\nu} \gamma^{\delta_{\nu,1}} \bar\psi_x \left[e^{\mu\delta_{\nu,1}} U_{x,\nu}\psi_{x+\hat\nu} - e^{-\mu\delta_{\nu,1}} U^\dagger_{x-\hat\nu,\nu}\psi_{x-\hat\nu}\right] + 2m \bar\psi_x\psi_x 
\right\} .
\end{align}
To derive a tensor-network formulation one changes to dual variables as prescribed in Refs.~\cite{Rossi:1984cv,Karsch:1988zx}. In the infinite-coupling limit, the \SU(3) gauge fields can be integrated out and one is left with an integral over Grassmann degrees of freedom. Because of the \SU(3) integration and the Grassmann nature of the remaining variables, the latter only contribute through mesonic combinations $\bar\psi_x\psi_x$
and baryonic (antibaryonic) combinations $B_x\,(\bar B_x)$ of 3 quarks (antiquarks).


\subsection{Tensor formulation}

To construct a tensor-network formulation, we now integrate out the mesonic contributions, but leave the baryonic contributions unaltered, as these would otherwise introduce \emph{non-local sign factors},
\begin{align}
Z = \sum_{\boldsymbol j}  \int \prod_x S^{(x)}_{\ind{j}} G^{(x)}_{\ind{l}}
\end{align}
with local numeric and Grassmann tensors:
\begin{align}
S^{(x)}_{\ind{j}} &= 
\delta_{x\in{\cal B}} \,  w_{\!\!\cal B}(\ind{l})
+ 
\delta_{x\in{\cal M}} \, w_{\!\!\cal M}(\ind{j})
\\
G^{(x)}_{\ind{l}} &=  (dB_x)^{\sum_\nu (l_{x,\nu}^{-}+l_{x,-\nu}^{+})} (d\bar B_x)^{\sum_\nu (l_{x,\nu}^{+}+l_{x,-\nu}^{-})} \prod_{\nu=1}^d  (B_{x} \bar B_{x+\hat\nu})^{l_{x,\nu}^-} (\bar B_x B_{x+\hat\nu})^{l_{x,\nu}^+} .
\end{align}
Each term in the partition function is characterized by its indices $\J=(j_{1,1},\ldots,j_{V,d})$,
where each link carries an index $0 \leq j_{x,\nu} \leq 5$ and the baryonic occupation numbers $l$ are determined by $j$, with $l\equiv l(j) \in\{-1,0,1\}$ and $l^{\pm} = \delta_{l,\pm1}$.
For the tensor indices we introduced the notation $\ind{j}\equiv j_{x,-1}j_{x,1} \dots j_{x,-d}j_{x,d}$, where $j_{x,-\nu} \equiv j_{x-\hat\nu,\nu}$.


\section{Grassmann HOTRG}

As the baryonic Grassmann variables are not integrated out, we use ideas of Grassmann tensor networks developed in GHOTRG \cite{Sakai:2017jwp} to perform the grid coarsening in the presence of the baryonic Grassmann variables.

First we decouple the $B$ and $\bar B$ Grassmann variables in the interaction terms by introducing one auxiliary Grassmann variable $c_{x,\nu}$ on \textit{each link} (recall $l_{x,\nu}^\pm \in \{0,1\}$):
\begin{align}
\begin{aligned}
(\bar B_{x}B_{x+\hat\nu})^{l_{x,\nu}^+} &= 
\left(\bar B_{x}B_{x+\hat\nu} \int dc_{x,\nu} c_{x,\nu}\right)^{l_{x,\nu}^+}  
= \int (\bar B_{x}c_{x,\nu})^{l_{x,\nu}^+} (B_{x+\hat\nu}dc_{x,\nu})^{l_{x,\nu}^+}, \\
(B_{x}\bar B_{x+\hat\nu})^{l_{x,\nu}^-} &= 
\left(B_{x}\bar B_{x+\hat\nu} \int d c_{x,\nu}  c_{x,\nu}\right)^{l_{x,\nu}^-} 
= \int (B_{x} c_{x,\nu} )^{l_{x,\nu}^-}(\bar B_{x+\hat\nu}dc_{x,\nu})^{l_{x,\nu}^-} .
\end{aligned}
\label{aux}
\end{align}
Note that the backward and forward baryon interactions on every link are mutually exclusive such that one auxiliary Grassmann variable per link is sufficient.

All factors in brackets in the right hand sides of \eqref{aux} are commuting and can be moved around freely in $Z$ to integrate out the (anti)baryons, without generating non-local sign factors.
After integrating out all $B_x$ and $\bar B_x$ the partition function becomes
\begin{align}
Z = \sum_{\boldsymbol j} \int \prod_x
T^{(x)}_{\ind{j}} 
K^{(x)}_{\ind{f}} ,
\end{align}
with Grassmann tensor
\begin{align}
K^{(x)}_{\ind{f}}  &= 
\prod_\nu(c_{x,\nu})^{\f_{x,\nu}} 
\revprod\nu(dc_{x,-\nu})^{\f_{x,-\nu}} \\
&= (c_{x,1})^{\f_{x,1}} \dots (c_{x,d})^{\f_{x,d}} (dc_{x,-d})^{\f_{x,-d}} \dots (dc_{x,-1})^{\f_{x,-1}} ,
\end{align}
where we defined the reverse ordered product $\inlinerevprod{\nu}$ and $c_{x,-\nu}\equiv c_{x-\hat\nu,\nu}$. 
Here, $f_{x,\nu}\equiv f_{x,\nu}(j_{x,\nu}) \in \{0,1\}$ is the \emph{Grassmann parity} of the corresponding index $j_{x,\nu}$.

The numeric tensor is
\begin{align}
T^{(x)}_{\ind{j}} 
&=
\omega_{\ind{l}} S^{(x)}_{\ind{j}} ,
\end{align}
with a local sign factor $\omega_{\ind{l}}$ coming from a rearrangement of the auxiliary Grassmann variables in $K^{(x)}$.
Note that the components of $T$ are nonzero only when the corresponding entry of $K$ is \emph{Grassmann even}.

\subsection{Blocking}

In GHOTRG the partition function is evaluated with an iterative blocking procedure, that consists of the following major steps: First, the Grassmann tensors are blocked such that the number of Grassmann variables is halved at each step. This generates a new local sign factor, which is absorbed in the coarse-grid numeric tensor. This tensor is then truncated using HOSVD, in the usual way of HOTRG, to keep the total tensor dimension under control.

\begin{figure}
\centerline{\includegraphics[scale=0.8]{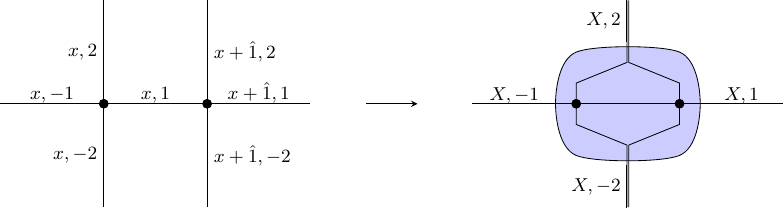}}
\caption{Illustration of a blocking in the 1-direction.
\label{fig:blocking}
}
\end{figure}

Let us consider a blocking in the 1-direction where two adjacent sites $(x,x+\hat1)$ are merged into a coarse lattice site $X$, as depicted in Fig.~\ref{fig:blocking},
\begin{align}
\sum_{\j_{x,1}}  \int_{c_{x,1}}
T^{(x)}_{\ind{j}} T^{(x+\hat1)}_{\indx{j}{x+\hat1}}
K^{(x)}_{\ind{j}}
K^{(x+\hat1)}_{\indx{j}{x+\hat1}} 
\quad\to\quad
\cR^{\xx} \cG^{\xx} .
\end{align}

\subsection{Grassmann blocking}

The first step is to block two adjacent Grassmann tensors and integrate out their shared link variable,
\begin{align}
\cG^{\xx} &\equiv 
\int_{c_{x,1}} K^{(x+\hat1)}_{\indx{f}{x+\hat1}}  K^{(x)}_{\ind{f}} .
\end{align}
This yields
\begin{align}
\lefteqn{\cG^{\xx}}
\notag\\
&= \int_{c_{x,1}} \! \prod_\nu(c_{x+\hat1,\nu})^{\f_{x+\hat1,\nu}}
\coprod_{\nu=2}^{d} (dc_{x+\hat1,-\nu})^{\f_{x+\hat1,-\nu}}
\underline{(dc_{x,1})^{\f_{x,1}} (c_{x,1})^{\f_{x,1}}}
\prod_{\nu=2}^d (c_{x,\nu})^{\f_{x,\nu}}
\revprod{\nu}(dc_{x,-\nu})^{\f_{x,-\nu}}
\notag\\
&= \sigma_{[\f_{x}\f_{x+\hat1}]} \, (c_{x+\hat1,1})^{\f_{x+\hat1,1}} 
\left[\prod_{\nu=2}^d (c_{x,\nu})^{\f_{x,\nu}}(c_{x+\hat1,\nu})^{\f_{x+\hat1,\nu}} \right] 
\left[\coprod_{\nu=2}^{d} (dc_{x+\hat1,-\nu})^{\f_{x+\hat1,-\nu}}(dc_{x,-\nu})^{\f_{x,-\nu}}\right]
(dc_{x,-1})^{\f_{x,-1}} .
\end{align}
For perpendicular directions ($\nu\geq 2$) the Grassmann variable $c_{x,\nu}$ and its differential $dc_{x,\nu}$ are not in same blocked tensor $\cG^{\xx}$, and moving the variables to the right position in the partition function to allow their integration would generate \emph{non-local sign factors}.

This difficulty can be overcome by introducing new, coarse-grid Grassmann variables $\tilde c_{X,\nu}$.
In practice, this is achieved by inserting an identity 
\begin{align}
\prod_{\nu=2}^d \left( \int d\cc_{X,-\nu} \cc_{X,-\nu} \right)^{\cf_{X,-\nu}} = 1
\end{align}
in every $\cG^{(\xx)}$, with
\begin{align}
\cf_{X,-\nu} \equiv (\f_{x,-\nu} + \f_{x+\hat1,-\nu}) \!\!\!\mod 2 .
\end{align}
If we then shift the \emph{commuting combinations}
\begin{align}
(\cc_{X,-\nu})^{\cf_{X,-\nu}}(dc_{x+\hat1,-\nu})^{\f_{x+\hat1,-\nu}}(dc_{x,-\nu})^{\f_{x,-\nu}} 
\end{align}
from the coarse site $X$ to $X-\hat\nu$, for all $\nu\geq2$ on the entire coarse lattice, then all pairs $(c_{x,\nu},c_{x+\hat 1,\nu})$  perpendicular to the contraction direction ($\nu\geq 2$) can be integrated out and are replaced by $\tilde c_{X,\nu}$ on the coarse lattice.
After blocking, the number of Grassmann variables in the partition function is effectively reduced from
$V d$ to $\frac12 V d$. Note that this is an exact procedure, which does not require any approximation.

The coarse grid partition function is now
\begin{align}
Z = \sum_{\boldsymbol j} \int \prod_{\X} \cS^{(\X)} \cK^{(\X)} 
\end{align}
with Grassmann tensor
\begin{align*}
\cK^{(\X)}
&= (c_{x+\hat1,1})^{\f_{x+\hat1,1}} 
\left[\prod_{\nu=2}^d (\cc_{X,\nu})^{\cf_{X,\nu}}\right]
\left[\coprod_{\nu=2}^d \left(d\cc_{X,-\nu}\right)^{\cf_{X,-\nu}} \right]
(dc_{x,-1})^{\f_{x,-1}} 
\end{align*}
\vspace{-1ex}
and numeric tensor
\begin{align*}
\cS^{(\X)}_{\j_{x,-1}\j_{x+\hat1,1}(\j_{x,-\nu},\j_{x+\hat1,-\nu})(\j_{x,\nu},\j_{x+\hat1,\nu})|_{\nu\neq 1}}
&= \sigma_{[\f_{x}\f_{x+\hat1}]}
\sum_{\j_{x,1}} 
T^{(x)}_{\ind{j}}
T^{(x+\hat1)}_{\indx{j}{x+\hat1}} 
\end{align*}
where a \emph{local sign factor} $\sigma_{[\f_{x}\f_{x+\hat1}]}$ was generated by reordering the Grassmann variables in $\cG^{(x,x+\hat1)}$ to perform these integrations. Note that the Grassmann parities $f\equiv f(j)$, and thus, the sign factor $\sigma$ does not generate new indices for the tensor $\cS^{(\X)}$.


\subsection{Truncating the numeric tensor}

After blocking the Grassmann tensors, we now perform a HOSVD of the coarse-grid numeric tensor $\cS^{(\X)}_{\j_{x,-1}\j_{x+\hat1,1}{(\j_{x,-\nu},\j_{x+\hat1,-\nu})(\j_{x,\nu},\j_{x+\hat1,\nu})|_{\nu\neq 1}}}$ to reduce the dimension of the perpendicular directions from $D^2 \to D$ and yield a truncated numeric tensor $\cT^{(\X)}_{\j_{x,-1}\j_{x+\hat1,1}{\cj_{\X,-\nu}\cj_{\X,\nu}|_{\nu\neq 1}}}$.

The numeric tensor has a particular Grassmann-parity structure as each nonzero entry can be attributed a definite Grassmann parity. After applying the HOSVD approximation, this property is conserved and the nonzero components of the truncated coarse-grid numeric tensor still have definite Grassmann parities, i.e., $\cf \equiv \cf(\cj)$ (for details see \cite{Bloch:2022vqz}). 

After one blocking step the shape of the (approximate) coarse-grid partition function is identical to that of the fine-grid partition function, albeit now a function of the coarse-grid indices and coarse-grid Grassmann variables. This property allows us to perform iterative blockings in any of the $d$ directions using the procedure described above.


\subsection{Blocking the complete lattice}

The same procedure can be repeated to contract any other direction. 
The blocking is iterated until the lattice is reduced to a single point. At this stage, the remaining Grassmann variables are integrated out after applying the boundary conditions, and the numeric tensor is traced to yield the partition function $Z$. Thermodynamical observables are computed using (stabilized) finite differences.

Let us make some brief remarks on the implementation.
When using the Grassmann-parity structure of the tensor, the cost of the GHOTRG method for sQCD is similar to that of HOTRG for purely numeric tensor networks. Without explicitly using this property, the cost of the algorithm would increase with a factor $2^{4d-1}$, i.e., $2^7$, $2^{11}$, $2^{15}$ in 2, 3, 4 dimensions, respectively.
The implementation of GHOTRG is slightly more complex than that of HOTRG, because the local sign factors have to be built-in during the contraction and truncation of the numeric tensors.


\section{Results for sQCD using GHOTRG in two dimensions}

\subsection{Validation}

\begin{figure}
\centering
\includegraphics[width=0.5\textwidth]{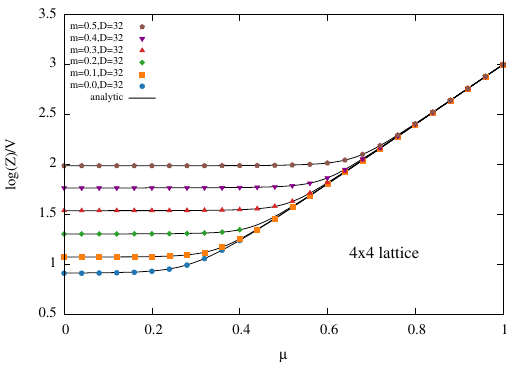}
\caption{Comparison of $\logZ/V$ computed using GHOTRG (with $D=32$) with the exact results, as a function of $\mu$ for various $m$.}
\label{fig:logZ}
\end{figure}

We first validated the GHOTRG method by computing the partition function on a $4\times4$ lattice and found very good agreement between the numerical tensor data and the exact results \cite{Bloch:2022vqz}. This agreement is illustrated in Fig.~\ref{fig:logZ}, where we show $\logZ/V$ as a function of $\mu$ and $m$.

\subsection{Chiral condensate and chiral symmetry}

\begin{figure}
\centering
\begin{minipage}[t]{0.48\textwidth}
\includegraphics[width=\textwidth]{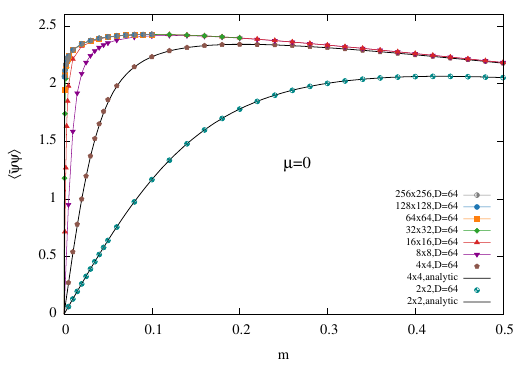}
\caption{Chiral condensate $\braket{\bar\psi\psi}$ as a function of the mass $m$ for various volumes $V$.}
\label{fig:ccm}
\end{minipage}
\hfill
\begin{minipage}[t]{0.48\textwidth}
\includegraphics[width=\textwidth]{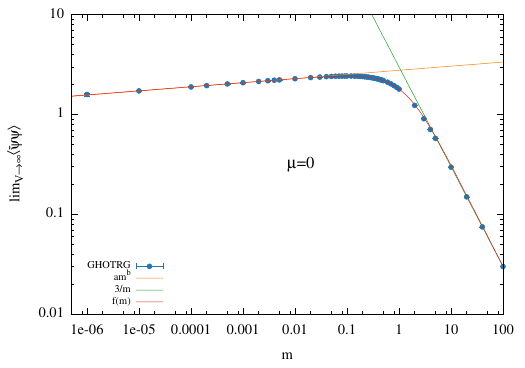}
\caption{Chiral condensate $\braket{\bar\psi\psi}$ for $D\to\infty$ as a function of the mass $m$ in the infinite-volume limit $V\to\infty$.}
\label{fig:dcsb}
\end{minipage}
\end{figure}

Next we investigate the chiral condensate
\begin{align}
\braket{\bar\psi\psi} = \frac{1}{V} \frac{\partial\ln Z}{\partial m}
\label{cc}
\end{align}
at zero chemical potential $\mu=0$. In Fig.~\ref{fig:ccm} we show the chiral condensate as a function of the mass $m$ for various volumes $V$ with bond dimension $D=64$. For fixed mass, the chiral condensate converges to its infinite volume limit as $V$ gets larger. As the mass gets smaller, larger volumes are needed to reach this limit. Note that the $2\times2$ and $4\times4$ data are compared with the exact results. 

The data for the chiral condensate can be used to investigate a possible dynamical breaking of the chiral symmetry, which would be signaled by
\begin{align}
\lim_{m\to0} \lim_{V\to\infty} \braket{\bar\psi\psi} \neq 0 .
\end{align}
To perform this investigation, we took the limit $D,V\to\infty$ for ever smaller masses. The results of these extrapolations are shown in Fig.~\ref{fig:dcsb}. The infinite-volume data can be well fitted by the empirical formula
\begin{align}
f(m) = \frac{a m^b +c m}{1+d m+(c/3)m^2}
\end{align}
with fitted parameter values $a = 2.77$, $b = 0.0409$, $c = 1.05$, $d = 0.770$, and
asymptotic behavior
\begin{align}
f(m) \sim 
\begin{cases}
3/m & \text{for large masses} ,
\\
a m^b & \text{for $m<0.005$} .
\end{cases}
\end{align}
The large-mass limit can immediately be read off from the partition function, but the small-mass behavior is a genuine dynamical property of the theory, which shows that the chiral symmetry is not dynamically broken in two-dimensional sQCD with (two tastes of) staggered quarks.


\subsection{Number density and chiral condensate versus at nonzero chemical potential}

\begin{figure}
\centering
\includegraphics[width=0.49\textwidth]{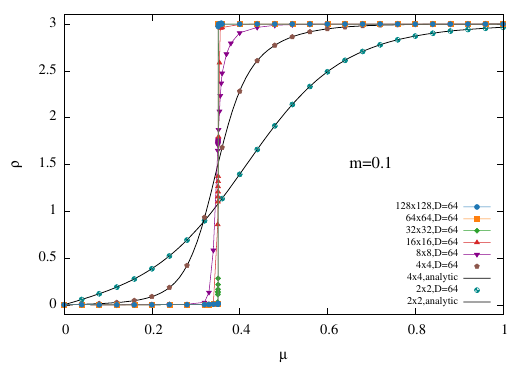}
\includegraphics[width=0.49\textwidth]{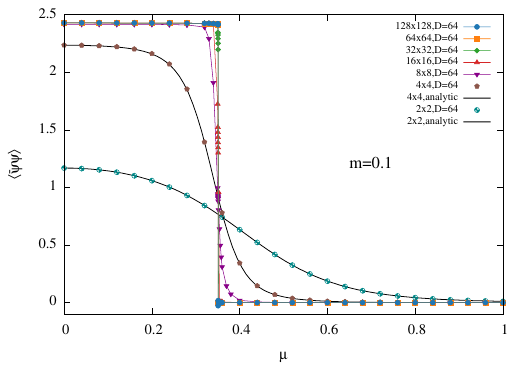}
\caption{Number density (left) and chiral condensate (left) as a function of the chemical potential $\mu$ and the volume for $m=0.1$ and $D=64$.}
\label{fig:nonzeromu}
\end{figure}

As mentioned in the introduction, one of the major reasons to use tensor-network methods is their ability to produce results for systems with complex actions. Therefore, we use the GHOTRG method to investigate  two-dimensional sQCD at nonzero chemical potential, where we compute the chiral condensate \eqref{cc} and the quark number density
\begin{align}
 \rho = \frac{1}{V} \frac{\partial \ln Z}{\partial \mu} .
\end{align}
In Fig.~\ref{fig:nonzeromu} we show the number density and the chiral condensate as a function of $\mu$ for $m=0.1$, computed with $D=64$. For the $2\times2$ and $4\times4$ lattices we also show the exact results. Clearly the large volume behavior shows a hint of a first-order phase transition, above which the chiral symmetry is effectively restored and the vacuum has a nonzero baryon density.

\subsection{Preliminary results in three dimensions}

\begin{figure}
\centering
\includegraphics[width=0.49\textwidth]{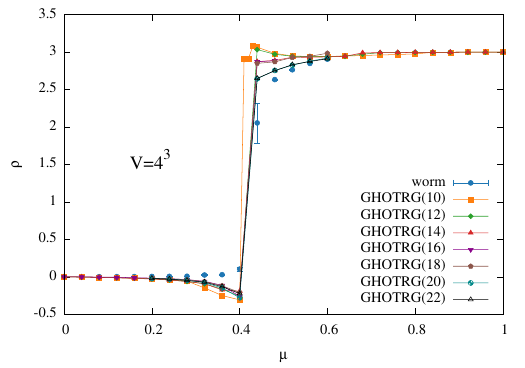}
\includegraphics[width=0.49\textwidth]{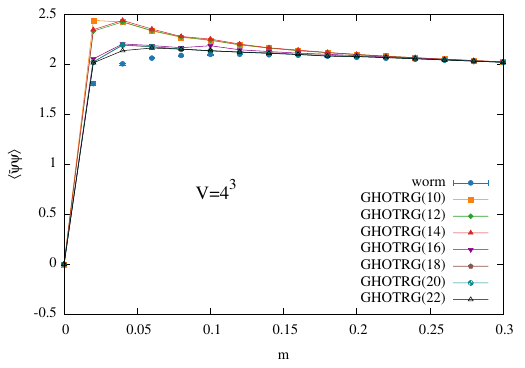}
\caption{Number density as a function of $\mu$ for $m=0$ (left) and
chiral condensate as a function of $m$ for $\mu=0$ (right) on a $4^3$ lattice for varying bond dimensions $D$, denoted by GHOTRG($D$) in the plots. The results are compared with those obtained with  the worm algorithm.}
\label{fig:3d}
\end{figure}

Finally, we present first GHOTRG results for sQCD in three dimensions, obtained on a $4^3$ lattice. In Fig.~\ref{fig:3d} we show the number density as a function of the chemical potential $\mu$ for $m=0$ and the chiral condensate as a function of $m$ for $\mu=0$, both on a $4^3$ lattice and using different bond dimensions $D$. The results are compared with those obtained with the worm algorithm \cite{Fromm:2010lga}. Although these preliminary results are encouraging, they clearly show the need to increase the bond dimension beyond $D>22$. In three dimensions, the HOTRG and GHOTRG methods scale as $D^{11}$. However, we have developed improved methods using additional hierarchical factorizations in three and four dimensions, which scale like $D^6$ and $D^8$, respectively. 
We are currently porting these methods to include the additional sign factors arising in the GHOTRG method, such that we should soon be able to get much better accuracies at a lower cost.


\section{Outlook}

In these proceedings we presented a Grassmann tensor-network method for sQCD, and applied the method to the two- and three-dimensional cases. We validated the method and showed that, in two dimensions, the chiral symmetry is not dynamically broken. Moreover, we found hints of a first order phase transition at some critical value of the quark chemical potential. Finally, some first GHOTRG results for three-dimensional sQCD were presented.

We are currently applying the GHOTRG method to sQCD in three and four dimensions. To achieve this goal we are extending our hierarchical HOTRG factorization scheme to GHOTRG, in order to reduce the scaling cost of the method to $D^6$ and $D^8$, respectively. We are also developing a GHOTRG tensor network to go beyond the infinite-coupling limit and include the next-to-leading order corrections in the $\beta$-expansion of $e^{S_G}$ with gauge action $S_G$.


\nocite{apsrev42Control}
\bibliographystyle{apsrev4-2.bst}
\bibliography{biblio,revtex-custom} 

\end{document}